\documentclass[journal]{IEEEtran}

\usepackage[T1]{fontenc}
\usepackage[utf8]{inputenc}
\usepackage{cite}
\usepackage{graphicx}
\usepackage[table]{xcolor}
\usepackage{svg}
\usepackage{booktabs}
\usepackage{tabularx}
\usepackage{array}
\usepackage{multirow}
\usepackage{url}
\usepackage{amsmath}
\usepackage{amssymb}
\usepackage{textcomp}
\usepackage{placeins}
\usepackage{flafter}
\usepackage{pifont}
\usepackage{needspace}
\usepackage{algorithm}
\usepackage{threeparttable}
\usepackage{siunitx}
\usepackage{algpseudocode}
\algrenewcommand\algorithmicrequire{\textbf{Input:}}
\algrenewcommand\algorithmicensure{\textbf{Output:}}
\newcommand{\findingbox}[1]{\vspace{0.4em}\noindent\fbox{\parbox{0.92\columnwidth}{#1}}\vspace{0.4em}}
\newcommand{\balancetail}[1]{\mbox{#1}}

\title{Heterogeneous Cross-Chain Transaction Tracing for Solana Bridges via Candidate-Set Selective Decision}
\author{Wenjie~Dou,
	Zheng~Che$^*$,
	Meng~Shen,
	Hanbiao~Du,
	Qing~Li,
	and~Yan~Qiang
	\thanks{Wenjie Dou, Zheng Che, Qing Li, and Yan Qiang are with the School of Software, North University of China, Taiyuan 030051, China (e-mail: s202513018@st.nuc.edu.cn; chezheng@nuc.edu.cn; 20230227@nuc.edu.cn; qiangyan@nuc.edu.cn).}
	\thanks{Meng Shen and Hanbiao Du are with the School of Cyberspace Science and Technology, Beijing Institute of Technology, Beijing 100081, China (e-mail: shenmeng@bit.edu.cn; duhanbiao@bit.edu.cn).}
	\thanks{$^*$Corresponding author: Zheng Che (chezheng@nuc.edu.cn).}%
}
\begin{document}
\bstctlcite{IEEEtranBSTCTL:dash}
\maketitle

\begin{abstract}
Solana is a rapidly growing high-throughput blockchain platform that has attracted substantial liquidity and user activity. However, this expansion has also drawn the attention of illicit actors, who frequently leverage cross-chain bridges to route illicit funds onto Solana to obfuscate transaction lineage. Unlike EVM-compatible platforms, Solana features distinct execution dynamics and lacks standard event logs, creating severe semantic gaps that prevent existing tracing methods from reliably correlating cross-ledger transactions. In this paper, we formalize four types of Solana-bound cross-chain transaction modes and propose a candidate-set selective decision-based tracing method called SolTracer. SolTracer maps disparate execution semantics into a unified event space and employs candidate-set selective decision-making to reliably associate target transactions while abstaining when valid targets are absent. Extensive experiments  demonstrate that SolTracer outperforms state-of-the-art (SOTA) methods across three representative scenarios: closed-world association, open-world association, and cross-source-chain generalization. In particular, under the challenging open-world setting with a 50\% TA ratio, SolTracer improves the F1 score by 20.16\% over the strongest SOTA baseline. Utilizing SolTracer,  we conduct an empirical analysis on real-world cross-chain transfers to investigate ecosystem dynamics. Our analysis explores the stark count-value divergence across bridge mechanisms, the prevalence of cross-asset shifts, and the decoupling between on-chain settlement and explorer visibility. 
\end{abstract}


\section{Introduction}

Driven by its ultra-low transaction fees and sub-second confirmation latency, Solana has rapidly emerged as a dominant high-throughput public platform. According to CoinMarketCap \cite{CoinMarketCap}, as of August 2026, Solana ranks among the top five cryptocurrencies globally by market capitalization. Furthermore, its daily decentralized exchange (DEX) trading volume has frequently surpassed that of the Ethereum, underscoring its pivotal role in the modern digital asset ecosystem.

\begin{figure}[!t]
	\centering
	\includegraphics[width=0.9\columnwidth]{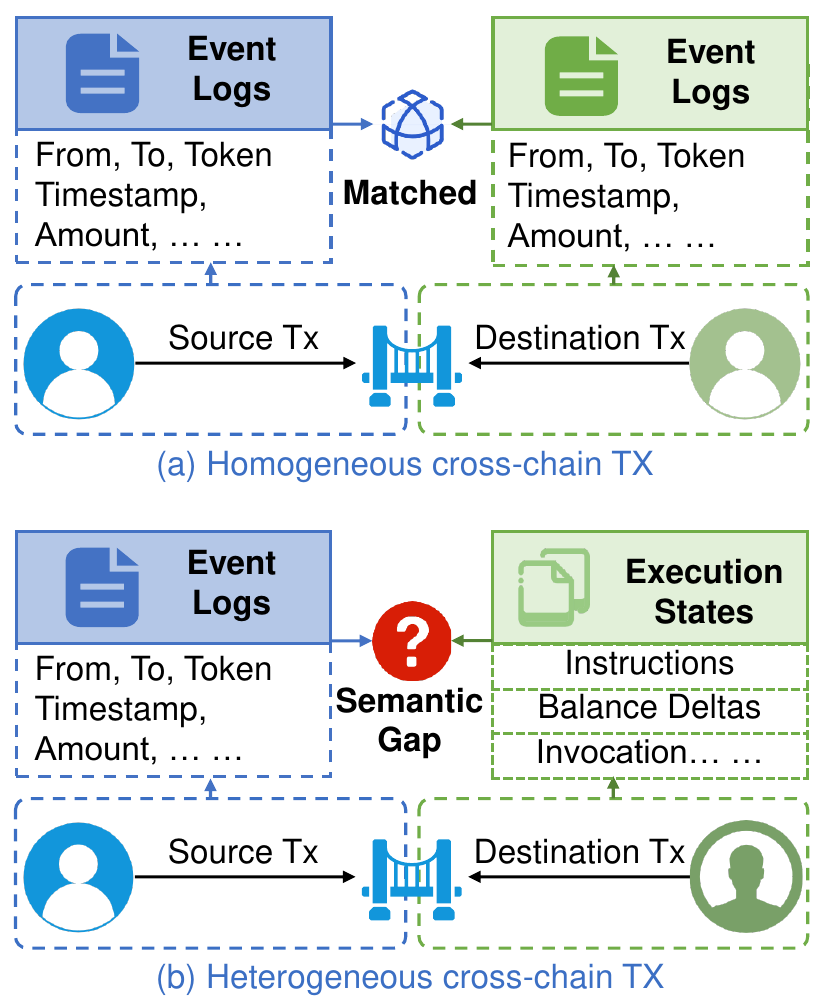}
	\caption{Comparison of transaction tracing between homogeneous and heterogeneous cross-chain transaction}
	\label{fig:1}
\end{figure}

Concurrently, the rapid proliferation and adoption of cross-chain bridge protocols have provided illicit actors with new avenues for money laundering and asset flight. Given Solana's massive liquidity and high throughput, cybercriminals frequently utilize its infrastructure to obfuscate transaction traces. For instance, in the Solana drainer campaigns targeting thousands of wallets, malicious actors exploited AllBridge to bridge over one million USD of stolen assets directly from Solana to Ethereum \cite{scamsniffer2024solana_drainers,decrypt2024solana_drainers}. Similarly, during the BingX exploit, attackers routed stolen funds onto Solana via deBridge, performed token swaps to disrupt asset lineage, and subsequently funneled them through multi-chain bridging services \cite{slowmist2024aml}.  Due to fundamental structural discrepancies between the source and target chains—such as heterogeneous account state models and execution logic—tracing transactions across these platforms becomes exceptionally difficult. As illustrated in Fig.~\ref{fig:1}, these architectural differences prevent heterogeneous cross-chain transactions from being easily correlated in the way homogeneous ones are.

Recent studies \cite{lin2025connector, lin2025tracktrace, liang2025connex, yousaf2019tracing, zhang2022cltracer, hu2024ice, yan2025evmpolygon} 
have investigated cross-chain transaction tracing primarily on Ethereum Virtual Machine-compatible (EVM-compatible) blockchains by correlating transaction attribute similarities or event logs between source and target chains. However, existing methods suffer from two fundamental challenges: \textbf{(1) Cross-Chain Heterogeneity.} Existing methods rely on uniform parsing schemes tailored for homogeneous receipts. In heterogeneous chains like Solana, the absence of standard event logs and the divergence of execution models lead to severe semantic gaps that prevent direct field alignment. \textbf{(2) Reliable Association.} Existing methods often assume that each source-chain transaction has a unique and directly identifiable counterpart on the target chain. This assumption does not always hold in real-world bridge scenarios, where asynchronous processing, cross-asset solver settlement, or unexecuted transfers may break the one-to-one correspondence between source and target transactions \cite{cao2026price}. Consequently, forcing such deterministic alignment can introduce substantial spurious associations and lead to a high false-positive rate.

In this paper, we focus on analyzing and tracing Solana-bound cross-chain transactions across heterogeneous ledgers. To address the semantic gap stemming from disparate blockchain execution models, we  first identify and summarize four types of Solana-bound cross-chain transaction modes based on their underlying settlement mechanisms.  To achieve reliable association across heterogeneous chains, we propose SolTracer, a cross-chain tracing method based on candidate-set selective decision, which maps disparate transaction semantics into a unified semantic event space and performs selective association over the candidate set through multi-field consistency verification. Finally, to gain deeper insights into the Solana cross-chain ecosystem, we conduct an empirical analysis to uncover critical security and observability characteristics within the Solana bridge ecosystem.

We summarize the main contributions as follows:
\begin{itemize}
    \item We identify and summarize four transaction modes for Solana-bound cross-chain transactions, i.e., Message Redemption, Pool Settlement, Solver Fulfillment, and Burn-Mint transactions. This abstraction formally characterizes the funding workflows and execution dynamics of various bridge mechanisms on Solana.
    \item We propose SolTracer, a cross-chain tracing method based on candidate-set selective decision, which maps disparate transaction semantics into a unified semantic event space and performs selective association over the candidate set through multi-field consistency verification.
    \item We evaluate the performance of SolTracer by comparing it to  six SOTA methods,  including  AttrMatch \cite{yousaf2019tracing},  EntityMatch \cite{yousaf2019tracing}, MultiMatch \cite{yan2025evmpolygon}, CONNECTOR~\cite{lin2025connector}, ABCTRACER~\cite{lin2025tracktrace} and ConneX~\cite{liang2025connex}.  SolTracer achieves the highest F1 scores in closed- and open-world settings.
    \item Based on the associated cross-chain trasactions, we conduct an empirical analysis to examine their count–value divergence, cross-asset shifts, target observability gaps, and mechanism-dependent association reliability, followed by a summary of several key findings.
\end{itemize}

The remainder of this paper is organized as follows. Section~\ref{sec: Background} reviews background and related work. Section~\ref{sec:  problem-definition} formalizes the cross-chain tracing problem, and Section~\ref{sec: tx-modes} defines the four transaction modes. Section~\ref{sec: SolTracer} presents SolTracer, followed by its experimental evaluation in Section~\ref{sec: Evaluation}. Section~\ref{sec: EmpiricalStudy} discusses empirical insights into the Solana bridge ecosystem, and Section~\ref{sec: conclusion} concludes the paper.

\FloatBarrier
\begin{figure}[!t]
\centering
\includegraphics[width=0.92\columnwidth]{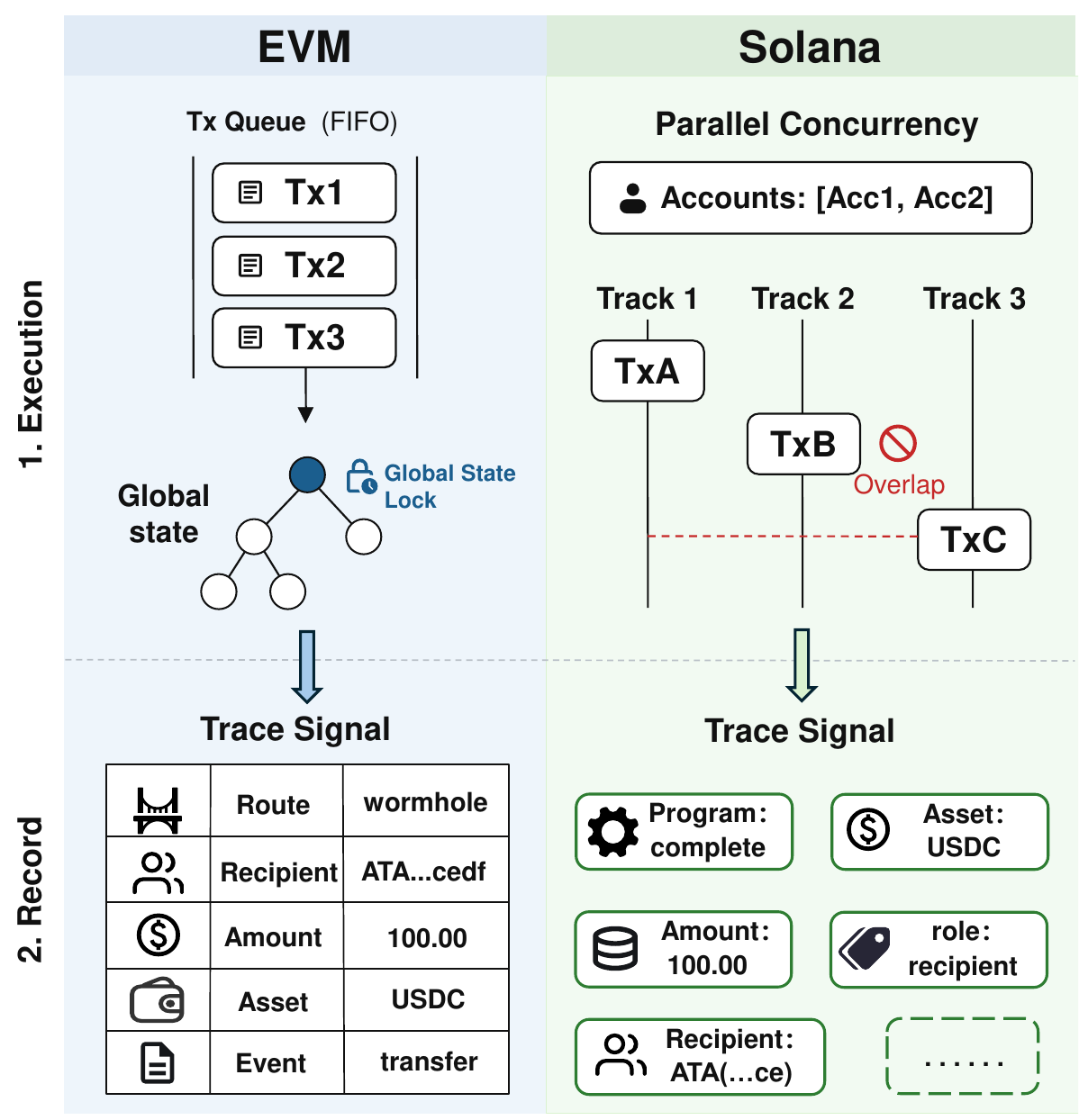}
\caption{Comparison of execution models and visible artifacts between EVM and Solana.}
\label{fig:different between solana and evm}
\end{figure}

\section{Background and Related Work}\label{sec: Background}
In this section, we first introduce the background of  cross-chain transactions and outline the key differences between EVM and Solana. We then review existing research on cross-chain transaction tracing.

\subsection{Cross-Chain Transactions}

Due to the lack of shared state across isolated blockchains, a cross-chain transaction fundamentally relies on the interaction among three key components: the source chain, the cross-chain bridge, and the target chain. 

\textit{1) Source Chain.} The cross-chain transfer begins on the source chain, where a user submits a transaction to record their transfer intent—such as the target chain, recipient, asset, and amount. Depending on the bridge mechanism, this step commits the initial state transition by locking, escrowing, or burning tokens, and emits event logs or state changes that serve as the origin record \cite{zhang2024bridges}.

\textit{2) Cross-Chain Bridge.} As the intermediate coordination layer, the bridge validates the validity of the source-chain transaction and relays the authorization to the destination network. Once the source transfer reaches finality, its verification mechanism verifies the state change through external validators, optimistic windows, or cryptographic proofs, and then issues a verified message, attestation, or proof for cross-ledger delivery \cite{lin2025tracktrace}.

\textit{3) Target Chain.} The final settlement occurs on the target chain, where a relayer or user submits the verified payload to the destination bridge contract. After validating the authorization and preventing replay, the contract executes the final state transition to deliver funds to the recipient, which includes actions such as releasing escrowed assets, minting tokens, or triggering local contract calls.

\begin{table*}[!t]
	\caption{Summary of Existing Studies on Cross-Chain Tracing}
	\label{tab:tracing_comparison}
	\centering
	\footnotesize
	\setlength{\tabcolsep}{3.5pt}
	\renewcommand{\arraystretch}{1.25}
	\begin{tabular}{@{}>{\centering\arraybackslash}m{2.2cm}>{\centering\arraybackslash}m{2.5cm}>{\centering\arraybackslash}m{2.8cm}>{\centering\arraybackslash}m{2.9cm}>{\centering\arraybackslash}m{3.0cm}>{\centering\arraybackslash}m{1.8cm}>{\centering\arraybackslash}m{1.8cm}@{}}
		\toprule
		\textbf{Category} & \textbf{Methods} & \textbf{Data Acquisition} & \textbf{Task} & \textbf{Methodology} & \textbf{Cross-Chain Heterogeneity} & \textbf{Reliable Association} \\
		\midrule
		\multirow{3}{*}{\shortstack{\textbf{Centralized}\\\textbf{Bridge Tracing}}}
		& Yousaf et al.~\cite{yousaf2019tracing} & ShapeShift API & Transaction association & Heuristic analysis & \textcolor{blue!70!black}{\ding{51}} & \textcolor{red!70!black}{\ding{55}} \\
		& CLTracer~\cite{zhang2022cltracer} & DE APIs & Account association & Heuristic analysis & \textcolor{blue!70!black}{\ding{51}} & \textcolor{red!70!black}{\ding{55}} \\
		& Hu et al.~\cite{hu2024ice} & Instant CE Service & Account association & Heuristic analysis & \textcolor{red!70!black}{\ding{55}} & \textcolor{red!70!black}{\ding{55}} \\
		\midrule
		\multirow{5}{*}{\shortstack{\textbf{Decentralized}\\\textbf{Bridge Tracing}}}
		& Yan et al.~\cite{yan2025evmpolygon} & Public Ledgers & Transaction association & Heuristic analysis & \textcolor{red!70!black}{\ding{55}} & \textcolor{red!70!black}{\ding{55}} \\
		& CONNECTOR~\cite{lin2025connector} & Public Ledgers & Transaction association & Machine learning & \textcolor{red!70!black}{\ding{55}} & \textcolor{red!70!black}{\ding{55}} \\
		& ABCTRACER~\cite{lin2025tracktrace} & RPC Services & Transaction association & Deep learning & \textcolor{red!70!black}{\ding{55}} & \textcolor{red!70!black}{\ding{55}} \\
		& ConneX~\cite{liang2025connex} & RPC Services & Transaction association & LLM and Heuristics & \textcolor{red!70!black}{\ding{55}} & \textcolor{red!70!black}{\ding{55}} \\
		\addlinespace[1.5pt]
		& \textbf{SolTracer} & \textbf{RPC Services \& APIs} & \textbf{Transaction association} & \textbf{Selective Decision} & \textcolor{blue!70!black}{\pmb{\checkmark}} & \textcolor{blue!70!black}{\pmb{\checkmark}} \\
		\bottomrule
	\end{tabular}
\end{table*}

\subsection{Differences Between EVM and Solana}

EVM and Solana exhibit fundamental differences in generating and exposing cross-chain trace evidence. We examine these differences from two primary perspectives: State Execution and Transaction Records, as summarized in Fig.~\ref{fig:different between solana and evm}.

\textit{1) Execution View.} The EVM applies smart-contract transactions as ordered updates to shared global state, whereas Solana organizes program execution around the accounts declared by each transaction. Solana can execute transactions concurrently when their account-access sets do not conflict \cite{SolPhish}, so its execution relation is constrained by account dependencies rather than a single global-state transition path.

\textit{2) Record View.} Transaction records are the ledger-visible data used to extract tracing evidence. While EVM smart contracts encapsulate bridge semantics within structured, ABI-decoded event logs embedded directly in transaction receipts, Solana fundamentally lacks such unified event logs and fragments bridge semantics across invoked programs, account roles, inner instructions, and balance deltas. Consequently, an observer cannot rely on a single receipt but must actively reconstruct and aggregate these disparate execution fragments before associating source- and target-side transactions.

\subsection{Related Work}
 In this section, we systematically review recent research on the cross-chain transaction tracing in cryptocurrencies.  We categorize existing research into two categories based on the bridge architecture: centralized bridge (CE bridge) tracing and decentralized bridge (DE bridge) tracing, as shown in Table~\ref{tab:tracing_comparison}.

\textit{1) CE bridge tracing.} Centralized bridges execute asset exchanges off-chain, exposing only decoupled deposit and withdrawal transactions on the public ledgers without intermediate matching traces. Hence, this type of research heavily depend on service-exposed APIs to collect operational data and apply heuristic rules over transaction metadata for association. Yousaf et al. \cite{yousaf2019tracing} combine ShapeShift records scraped from a public API with transactions from eight blockchains and use temporal and value heuristics to recover deposit--withdrawal flows. Zhang et al. \cite{zhang2022cltracer} replaces real-time monitoring with account-relationship-based historical transaction discovery and cross-ledger clustering. Hu et al. \cite{hu2024ice} collect instant exchange services through aggregator sites and controlled user interactions, then correlate deposits and withdrawals using temporal and pricing constraints. 

\textit{2) DE bridge tracing.} Decentralized bridges execute transfers through on-chain smart contracts, providing verifiable and auditable ledger records. Hence, this type of research mainly exploit these transparent contract traces, event logs, and transfer values to reconstruct cross-chain pairs. Yan et al. \cite{yan2025evmpolygon} associate Ethereum--Polygon transfers using shared user accounts alongside temporal, value, and token constraints. Lin et al. \cite{lin2025connector} identify source-side deposits from contract traces and correlate target-side withdrawals via bridge execution logs. To uncover less obvious relationships, Lin et~al.~\cite{lin2025tracktrace} apply named-entity recognition and information retrieval over event logs to capture implicit cross-chain cues.More recently, Liang et~al.~\cite{liang2025connex} leverage large language models for automatic semantic key--value extraction, validating candidate associations via value checks. 

\textit{3) Summary.} The limitations of existing methods are primarily reflected in two aspects. First, existing methods rely on homogeneous ledger assumptions, extracting explicit cues primarily from ABI-decoded event logs and structured transaction receipts. These parsers fail on heterogeneous ecosystems like Solana, where standard event logs are absent and execution semantics are scattered across accounts, inner instructions, and balance deltas. Second, current association heuristics enforce a rigid one-to-one closed-world mapping without a candidate-set-level reject option. Consequently, when target transactions are absent due to delayed relaying, solver batching, or unexecuted transfers, existing ranking schemes inevitably select the top-scoring distractor and produce false associations.

\section{Problem Definition}\label{sec: problem-definition}

This paper focuses on tracing Solana-bound cross-chain transactions by addressing target-transaction observability gaps across heterogeneous execution models. 
Formally, given an observed source-side bridge-out transaction $o_s \in \mathcal{O}_S$ on an EVM-compatible ledger, a protocol-compatible filtering mechanism first extracts a bounded candidate set $\mathcal{C}(o_s) = \{o_{t,1}, o_{t,2}, \ldots, o_{t,k}\} \subseteq \mathcal{O}_T$, where $\mathcal{O}_T$ denotes the Solana target observation space. 
Let $o^* \in \mathcal{O}_T \cup \{\bot\}$ represent the ground-truth Solana target transaction corresponding to $o_s$, where $o^* = \bot$ indicates that the cross-chain execution has not settled on-chain or falls beyond the observation scope. 
Consequently, the candidate set is \emph{not} guaranteed to contain a valid target (i.e., $o^* \notin \mathcal{C}(o_s)$ may occur) due to delayed relaying, solver batching, or unexecuted transfers.

To achieve reliable tracing under target uncertainty, we model the association process via a selective decision function $F(o_s, \mathcal{C}(o_s)) \in \mathcal{C}(o_s) \cup \{\bot\}$, defined as:
\begin{equation}
	F(o_s, \mathcal{C}(o_s)) = 
	\begin{cases}
		\hat{o}_t, & \text{if } \Gamma(o_s, \hat{o}_t \mid \mathcal{C}(o_s)) = 1, \\
		\bot, & \text{otherwise,}
	\end{cases}
	\label{eq:selective-decision}
\end{equation}
where $\hat{o}_t = \arg\max_{o_t \in \mathcal{C}(o_s)} \mathcal{S}(o_s, o_t)$ is the top-ranked candidate evaluated by a heterogeneous semantic scoring function $\mathcal{S}(\cdot)$, and $\Gamma(\cdot) \in \{0, 1\}$ is a selective decision rule that accepts the candidate only when its association evidence significantly dominates competing distractors. 

Based on this formulation, the design goals of our tracing system are twofold:

\textbf{1) Heterogeneous Semantic Association:} 
When the ground-truth target is present in the candidate set ($o^* \in \mathcal{C}(o_s)$), the objective is to reconstruct unified semantic features across disparate execution models such that the true target achieves the dominant matching score, yielding $\hat{o}_t = o^*$ and $F(o_s, \mathcal{C}(o_s)) = o^*$.

\textbf{2) Reliable Association:} 
When a reliable association cannot be guaranteed—either because the ground-truth target is absent ($o^* \notin \mathcal{C}(o_s)$) or the candidate set exhibits high semantic ambiguity—the decision rule should reject all candidates ($\Gamma(\cdot) = 0$) and output $\bot$, thereby provably preventing false-positive attributions.

\section{Solana-Bound Cross-Chain Transaction Modes}\label{sec: tx-modes}

\begin{figure}[h]
	\centering
	\includegraphics[width=0.5\textwidth]{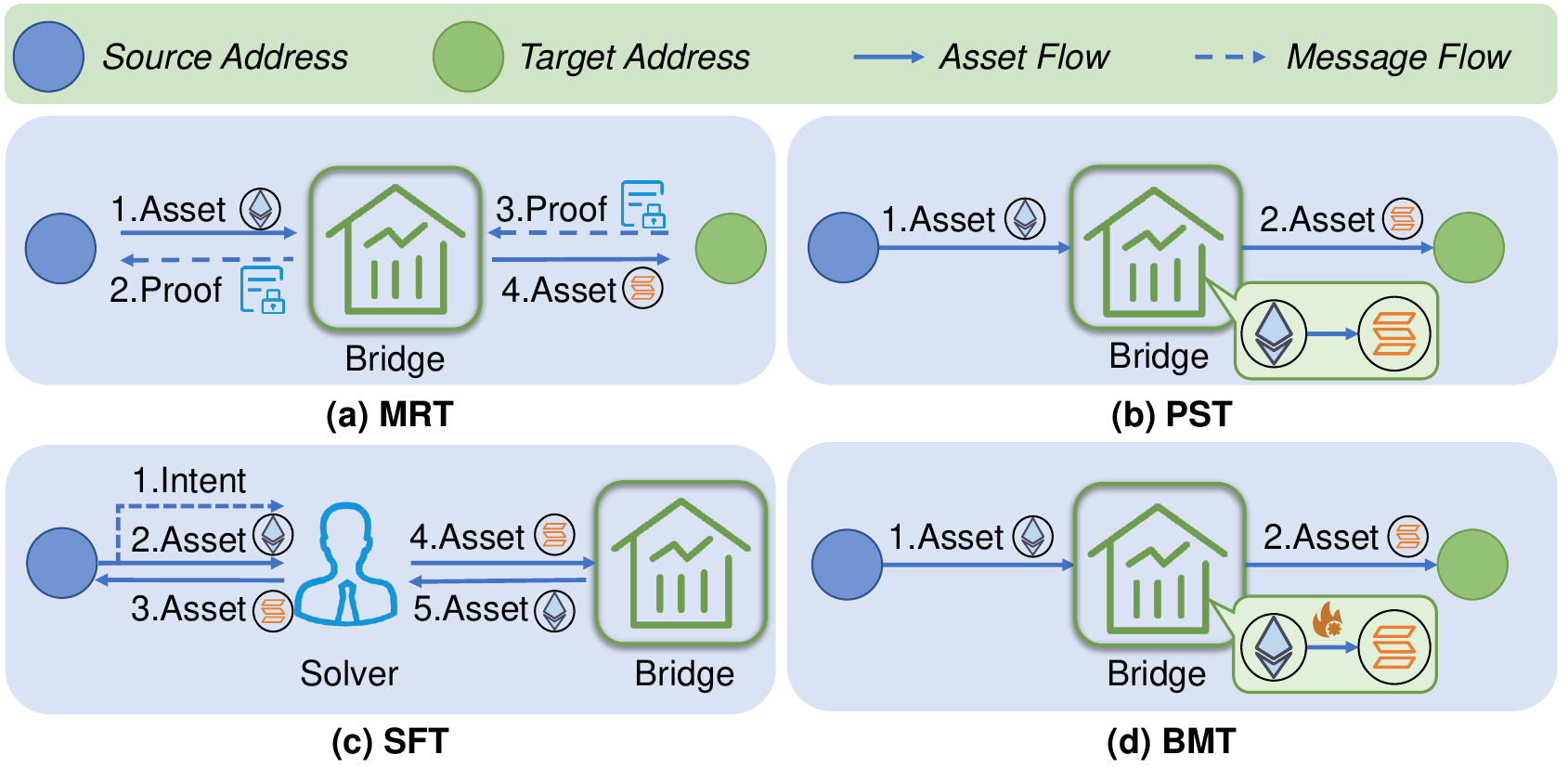}
	\caption{Solana-Bound Cross-Chain Transaction Modes}
	\label{fig:transaction-modes}
\end{figure}

In this section, we formalizes four Solana-bound transaction modes based on underlying funding topologies: Message Redemption Transactions, Pool Settlement Transactions, Solver Fulfillment Transactions, and Burn–Mint Transactions. For each mode, we characterize its execution workflow and analyze the observability gaps that hinder cross-chain correlation. The four transaction modes are illustrated in Fig.~\ref{fig:transaction-modes}.

\subsection{Message-Redemption Transactions, MRT}

MRT operates on a message-verification workflow where a verified source-chain attestation is redeemed on Solana to trigger a target-side release or minting operation (Fig.~\ref{fig:transaction-modes}(a)). Representative implementations include Wormhole \cite{wormhole-api} and deBridge \cite{debridge-api}. Under this mode, target-side evidence generally preserves source asset and amount semantics following asset mapping and decimal normalization. Due to Solana's account architecture, the recipient is explicitly designated through a program-derived Associated Token Account (ATA) on the target chain.

The main observability gaps arise from:

\begin{itemize}
\item \textbf{Account Transformation}. Heterogeneous account formats prevent direct comparison between the source recipient and the corresponding Solana token account.

\item \textbf{Indirect CPI Visibility}. Bridge programs transfer assets via Cross-Program Invocations (CPIs) to the Solana Program Library (SPL) Token program. Top-level instruction parsers only observe the root bridge invocation, requiring manual aggregation across inner instructions and balance deltas to recover the recipient and transferred amount.
\end{itemize}

\subsection{Pool-Settlement Transactions, PST}
PST executes cross-chain transfers by drawing liquidity from a shared Solana-side pool to pay the recipient (Fig.~\ref{fig:transaction-modes}(b)). Representative implementations include Allbridge Core  \cite{allbridge-api}and  Stargate V1 \cite{stargate_architecture}. The target payout amount frequently diverges from the initial source amount due to protocol fees, dynamic slippage, pool liquidity states, or intermediate conversions, resulting in route-dependent amount relations.

The main observability gaps arise from:

\begin{itemize}
\item \textbf{Amount Variation}. Fees, slippage, and route-specific conversion can change the target payout, invalidating exact comparison with the source amount.

\item \textbf{Shared-Pool Ambiguity}. A pool may serve multiple transfers within the same observation window, producing several payouts that all fall within the same time window and amount tolerance.
\end{itemize}

\subsection{Solver-Fulfillment Transactions, SFT}
SFT operates under an intent-based paradigm where the source-chain transaction records a user's target parameters, including the destination chain, recipient, requested asset, and minimum acceptable output (Fig.~\ref{fig:transaction-modes}(c)). Independent third-party solvers or relayers fulfill these conditions on Solana directly using proprietary liquidity, as exemplified by deBridge DLN, Mayan Swift, and Across \cite{debridge_dln_overview,mayan_swift_overview,across_intent_lifecycle}. Consequently, the target settlement forms a decoupled execution rather than an on-chain continuation of the source fund flow.

The main observability gaps arise from:

\begin{itemize}
\item \textbf{Cross-Asset Fulfillment}. Under cross-asset routes, the delivered asset may differ from the source asset, making raw asset equality an unreliable association signal.

\item \textbf{Funding-Path Separation}. Funds come from a solver-controlled account rather than a bridge-controlled account. This removes direct asset-flow continuity and weakens account-based links, although intent or solver evidence may remain visible.
\end{itemize}

\subsection{Burn-Mint Transactions, BMT}
BMT completes cross-chain transfers by burning canonical tokens on the origin ledger and minting native target tokens on Solana upon presenting a cryptographically verified off-chain attestation (Fig.~\ref{fig:transaction-modes}(d)). Representative implementations include Circle CCTP \cite{circle-api} and LayerZero Solana OFT routes \cite{layerzero_solana_oft}. While recipient identities and normalized values are generally preserved across chains, the authorizing attestation logic and the final minting execution are split across distinct transaction phases and data layers.

The main observability gaps arise from:

\begin{itemize}
\item \textbf{Authorization-Evidence Separation}. The attestation is produced off chain, while its payload and verification effects may appear on chain. Ledger-only analysis can therefore observe the mint without recovering the full provenance that authorizes it.

\item \textbf{Mint-Candidate Ambiguity}. Several mint operations with similar asset, amount, and timing attributes may occur within the same observation window, making time and value evidence insufficient to identify the intended target among competing mint candidates.
\end{itemize}

Across the four modes, heterogeneous execution breaks source-to-target semantic continuity in four dimensions: recipient account representations, transferred amount dynamics, asset identity consistency, and authorization provenance. Consequently, existing heuristics that rely on static attribute alignments fail to generalize across heterogeneous bridges \cite{yousaf2019tracing,yan2025evmpolygon}. Bridging these semantic gaps requires decoding Solana-specific CPIs, account roles, and off-chain attestations—concepts that cannot be captured by uniform EVM-tailored heuristics—which motivates the design of SolTracer.

\section{The Proposed SolTracer}\label{sec: SolTracer}

In this section, we propose SolTracer, a candidate-set selective decision method designed for tracing Solana-bound cross-chain transactions across heterogeneous blockchains. 

\subsection{Overview of SolTracer}

\begin{figure*}[!t]
\centering
\includegraphics[width=\textwidth]{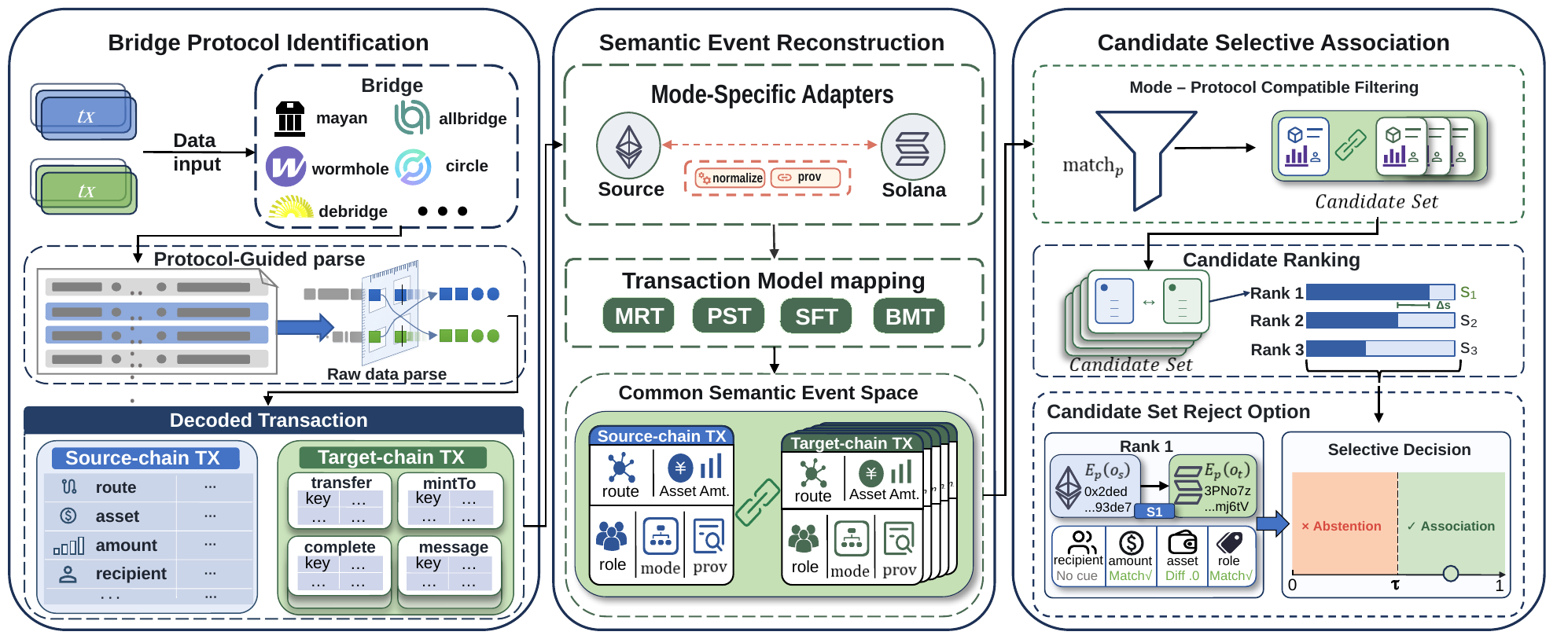}
\caption{System overview of SolTracer, which consists of three modules, i.e., Bridge Protocol Identification, Semantic Event Reconstruction, and  Candidate Selective Association.}
\label{fig:SolTracer}
\end{figure*}

The main components of SolTracer are threefold, namely the bridge protocol identification module, the semantic event reconstruction module, and the candidate selective association module. The overview of the SolTracer is shown in Fig.\ref{fig:SolTracer}.

\textit{1)  Bridge Protocol Identification.} This module identifies the underlying bridge protocol from source-chain routing cues and decodes the initial cross-chain intent, including route direction, asset mappings, and participant parameters. It also retrieves protocol-compatible transaction fragments from the Solana while admitting only supported transfers, thereby filtering out incompatible observations prior to candidate construction.

\textit{2) Semantic Event Reconstruction.} This module maps the disparate execution fragments of accepted transactions into a unified semantic event schema via mode-specific adapters. It systematically aggregates EVM receipts with Solana's inner instructions, account roles, and balance deltas across the four transaction modes (MRT, PST, SFT, and BMT), producing standardized semantic events with full field provenance to serve as inputs for the candidate selective association module.

\textit{3)  Candidate Selective Association.} This module constructs a bounded candidate set for each source intent and scores candidate pairs using multi-field semantic consistency across time, asset, amount, and participant roles. It further incorporates a candidate-set-level reject option that dynamically assesses candidate separation, thereby achieving reliable association when the true target exists while provably abstaining from false attributions when the target is absent.

\begin{algorithm}[!b]
	\caption{Bridge Protocol Identification}
	\label{alg:BridgeIdentification}
	\small
	\begin{algorithmic}[1]
		\Require Source observation $o_s$, protocol specifications $\Pi$, threshold $\tau_{\mathrm{proto}}$, observation pool $\mathcal{O}_T$
		\Ensure Decoded evidence tuple $(p, d_s, \mathcal{D}_p)$ or selective abstention $\bot$
		\State $p^* \gets \arg\max_{q \in \Pi} S_q(o_s)$ 
		\If{$S_{p^*}(o_s) < \tau_{\mathrm{proto}}$}
		\State \Return \Call{Abstain}{out-of-support}
		\EndIf
		\State $p \gets p^*$
		\State $d_s \gets \Call{DecodeSource}{o_s, p}$ 
		\State $\mathcal{O}_p \gets \Call{RetrieveTarget}{\mathcal{O}_T, p}$
		\State $\mathcal{D}_p \gets \Call{DecodeTargetFragments}{\mathcal{O}_p, p}$ 
		\State \Return $(p, d_s, \mathcal{D}_p)$
	\end{algorithmic}
\end{algorithm}

\subsection{Bridge Protocol Identification}

This module receives $o_s$ and the Solana observation pool $\mathcal{O}_T$. Its primary objective is to determine whether the source observation is supported by the protocol knowledge base available to SolTracer and, if so, to decode the source intent parameters and retrieve protocol-compatible Solana evidence required by downstream modules.

The protocol registry serves as a formal knowledge layer that can be extended via expert-guided specifications and semi-automated learning from protocol documentation and verified cross-chain transactions. Each protocol specification encapsulates stable invariants, including router and program identifiers, event topics, route directions, asset mappings, expected latency windows, amount transfer semantics, and participant account roles, without caching ground-truth target transactions.

Formally, let $\Pi$ denote the set of candidate protocol specifications. For each protocol $q \in \Pi$, SolTracer evaluates the rule-level compatibility of the observed source transaction $o_s$:
\begin{equation}
	S_q(o_s) = \sum_{j=1}^{K} w_j \mathbb{I}\!\left[\phi_j(o_s) \models r_{q,j}\right],
	\label{eq:protocol-score}
\end{equation}
where $\phi_j(o_s)$ denotes the $j$-th observed routing cue, $r_{q,j}$ is the corresponding matching rule for protocol $q$, and $w_j$ represents its assigned weight. SolTracer then identifies the best-matching protocol candidate:
\begin{equation}
	p = \underset{q \in \Pi}{\arg\max}\; S_q(o_s).
	\label{eq:protocol-selection}
\end{equation}
The identified protocol $p$ is accepted if and only if $S_p(o_s) \ge \tau_{\mathrm{proto}}$. Otherwise, the module outputs the abstention symbol $\bot$ to signify an unsupported transfer. Algorithm~\ref{alg:BridgeIdentification} summarizes this protocol identification and evidence retrieval workflow.

Through this mechanism, the module decodes key source intent parameters alongside compatible Solana execution traces (such as program IDs, account roles, and balance deltas). By enforcing protocol-level admission control prior to candidate-set construction, SolTracer prunes unsupported or malformed transfers at an early stage, preventing incompatible noise from polluting candidate sets and mitigating false associations induced by forced matching.

\subsection{Semantic Event Reconstruction}
The identified protocol $p$ determines the execution route and activates the corresponding mode adapter. For each observation $o$, the adapter integrates raw execution traces $\text{Exec}(o)$ and state transitions $\text{State}(o)$ into a unified semantic event:
\begin{equation}
	\begin{split}
		E_p(o) &= \text{Map}_p(o, \text{Exec}(o), \text{State}(o)) \\
		&= \langle \text{id}, t, \mathcal{R}, \mathcal{P}, \mathcal{A}, \mathcal{V}, \Phi \rangle,
	\end{split}
\end{equation}
where $\text{Exec}(o)$ captures transaction receipts, event logs, inner CPI trees, and program logs. $\text{State}(o)$ records balance transitions ($\Delta B = \text{postBalances} - \text{preBalances}$); $\mathcal{R}$, $\mathcal{P}$, $\mathcal{A}$, and $\mathcal{V}$ denote the route domain, participant roles, normalized asset identifier, and decimal-adjusted amount, respectively. $\Phi$ stores provenance metadata tracking underlying evidence layers.

\begin{algorithm}[t]
	\small 
	\caption{Semantic Event Reconstruction}
	\label{alg:semantic_reconstruction}
	\begin{algorithmic}[1]
		\Require Obs. $o$, proto. $p$, traces $\text{Exec}(o)$, state $\text{State}(o)$
		\Ensure Semantic event $E_p(o) = \langle \text{id}, t, \mathcal{R}, \mathcal{P}, \mathcal{A}, \mathcal{V}, \Phi \rangle$
		\State $E \leftarrow \text{InitHeader}(o)$, $\Phi \leftarrow \emptyset$, $m \leftarrow \text{GetMode}(p)$
		\If{$m = \text{\textbf{MRT}}$}
		\State $(\mathcal{P}, \mathcal{A}, \mathcal{V}) \leftarrow \text{ExtractCPI\_ATA}(\text{Exec}(o), \text{State}(o))$
		\State $\Phi \leftarrow \{\text{CPI}, \text{ATA}\}$
		\ElsIf{$m = \text{\textbf{PST}}$}
		\State $(\mathcal{P}, \mathcal{A}, \mathcal{V}) \leftarrow \text{ExtractPoolDelta}(\text{State}(o), p)$
		\State $\Phi \leftarrow \{\text{BalanceDelta}, \text{Fee}\}$
		\ElsIf{$m = \text{\textbf{SFT}}$}
		\State $(\mathcal{P}, \mathcal{A}, \mathcal{V}) \leftarrow \text{ExtractSolverPayout}(\text{Exec}(o), \text{State}(o))$
		\State $\Phi \leftarrow \{\text{OrderIntent}, \text{SolverTrace}\}$
		\ElsIf{$m = \text{\textbf{BMT}}$}
		\State $(\mathcal{P}, \mathcal{A}, \mathcal{V}) \leftarrow \text{ExtractMintProof}(\text{Exec}(o))$
		\State $\Phi \leftarrow \{\text{Attestation}, \text{MintCPI}\}$
		\EndIf
		\State \Return $E \cup \langle \mathcal{P}, \mathcal{A}, \mathcal{V}, \Phi \rangle$
	\end{algorithmic}
\end{algorithm}

Algorithm~\ref{alg:semantic_reconstruction} implements the semantic reconstruction across disparate transaction modes, projecting mode-specific execution fragments into a common semantic space.  For MRT, the adapter unrolls inner CPI trees directed to the SPL Token program, resolves the program-derived ATA to the recipient, and normalizes decimal values. For PST, it extracts net balance deltas from $\text{State}(o)$ to absorb pool fees and dynamic slippage, bounding the payout amount within route-specific intervals. For SFT, it parses intent-fulfillment logs to decouple intermediate solver funding accounts from actual recipients, accommodating cross-asset swaps.  For BMT, it reconciles off-chain attestation proofs with on-chain \texttt{mintTo} invocations across split execution phases. 
All adapters output the uniform schema $E_p(o)$ for downstream candidate selective association.

\subsection{Candidate Selective Association}\label{subsec: Selective Association}

The third module uses the common events to build the candidate set defined in Section~\ref{sec: problem-definition}. Given $o_s$ and the accepted protocol $p$, SolTracer applies protocol-compatible filtering to the Solana observation pool:
\begin{equation}
\mathcal{C}(o_s)=
\left\{o_{t,i}\in\mathcal{O}_T\;\middle|\;
\operatorname{match}_p\!\left(E_p(o_s),E_p(o_{t,i})\right)=1
\right\},
\label{eq:candidate-set}
\end{equation}
where $\operatorname{match}_p$ checks the route and time window, mapped or expected target asset, mode-specific amount relation, and recovered participant roles. It therefore supports fee-adjusted PST payouts and cross-asset SFT fulfillment without enforcing strict equality over raw assets and amounts. An empty set indicates candidate-space \balancetail{abstention rather than a ranking failure.}

Let $\mathcal{C}$ denote $\mathcal{C}(o_s)$ in the rest of this subsection. For every candidate pair retained after filtering, SolTracer compares the reconstructed events by time distance, normalized or expected amount, direct or mapped asset agreement, participant roles, mode-specific evidence, and provenance completeness. A pairwise scorer ranks the candidates as $o_{t,(1)},o_{t,(2)},\ldots$. Ranking alone is insufficient because every nonempty candidate set has a top-ranked element. SolTracer therefore applies a lightweight candidate-set-level reject option. Let $\widetilde{\mathbf{z}}(\mathcal{C})$ denote the standardized candidate-set vector containing the top score, score separation, time and amount consistency, and asset, recipient, and semantic-role consistency. A logistic model estimates whether the candidate set supports a valid association:
\begin{equation}
P_{\mathrm{set}}(\mathcal{C})=
\sigma\!\left(\mathbf{w}^{\mathsf T}
\widetilde{\mathbf{z}}(\mathcal{C})+b\right),
\label{eq:candidate-set-support}
\end{equation}
where $\mathbf{w}$ and $b$ are learned jointly from target-present and verified target-absent candidate sets using class-balanced training. The operating threshold $\tau$ is selected on calibration data.

This candidate-set score implements the reliability rule in Eq.~\eqref{eq:selective-decision} via $\Gamma(o_s, \hat{o}_t \mid \mathcal{C}) = \mathbb{I}[P_{\mathrm{set}}(\mathcal{C}) \ge \tau]$. When $\mathcal{C} \neq \emptyset$ and $P_{\mathrm{set}}(\mathcal{C})\ge\tau$, the decision function $F$ returns the top-ranked candidate $\hat{o}_t$. Otherwise, it returns $\bot$. The recorded reason distinguishes candidate-space abstention, where $\mathcal{C}$ is empty, from candidate-set-level abstention, where a nonempty set fails to reach $\tau$. Consequently, selective association prevents forced false-positive attributions without changing within-set ranking.

\section{Performance Evaluation}\label{sec: Evaluation}
In this section, we conduct extensive experiments to evaluate the effectiveness of SolTracer. We first construct the dataset by exploiting real-world EVM-to-Solana cross-chain records across representative bridge protocols.  Secondly,  we introduce the implementation details, followed by a sensitivity analysis. Thirdly, we thoroughly compare SolTracer with the SOTA methods across three typical scenarios: closed-world,  open-world and generalization. Finally, we perform ablation studies to assess the contributions of each module within SolTracer.

\subsection{Dataset Construction}\label{sec:dataset}

We collect and verify real-world Solana-bound cross-chain transfers
from four EVM-compatible source chains, including Ethereum, Arbitrum,
Base, and Polygon PoS, across four representative bridges covering all
four transaction modes. Across the main and generalization
data-collection rounds, our acquisition pipelines retrieved 38,317
bridge records and retained 7,082 verified source--target pairs after
deduplication, dual-chain verification, and bridge-specific consistency
checks.

\begin{table}[!t] 
	\caption{Summary of the Collected and Labeled Dataset.} 
	\label{tab:bridge_datasets} 
	\centering 
	\scriptsize 
	\setlength{\tabcolsep}{2pt} 
	\renewcommand{\arraystretch}{1.10} 
	\resizebox{\columnwidth}{!}{%
		\begin{tabular}{@{}lccccc@{}} 
			\toprule 
			\multirow{2}{*}{Bridge}
			& \multirow{2}{*}{Raw Data}
			& Ethereum
			& Arbitrum
			& Base
			& Polygon \\ 
			& 
			& (CP/NCP)
			& (CP/NCP)
			& (CP/NCP)
			& (CP/NCP) \\ 
			\midrule 
			
			Wormhole 
			& 5,977 
			& 1,082/32,460 
			& 68/2,040 
			& 279/8,370 
			& 196/5,880 \\ 
 
			Allbridge
			& 1,480 
			& 795/23,850 
			& 182/5,460 
			& 182/5,460 
			& 182/5,460 \\ 
 
			deBridge 
			& 1,600 
			& 997/29,910 
			& 200/6,000 
			& 200/6,000 
			& 200/6,000 \\ 
 
			Circle 
			& 29,260 
			& 1,495/44,850 
			& 385/11,550 
			& 477/14,310 
			& 162/4,860 \\ 
			
			\midrule 
			Total 
			& 38,317 
			& 4,369/131,070 
			& 835/25,050 
			& 1,138/34,140 
			& 740/22,200 \\ 
			\bottomrule 
		\end{tabular}%
	} 
\end{table}

\begin{itemize}
\item \textbf{Wormhole:}
We use Wormhole WTT as the representative MRT product and collect
5,977 transfer records from WormholeScan. After cross-file
deduplication and dual-chain RPC verification, 1,625 valid
source--target pairs are retained, including 1,082 from Ethereum,
68 from Arbitrum, 279 from Base, and 196 from Polygon PoS.

\item \textbf{Allbridge:}
We use Allbridge Core as the representative PST product and collect
1,480 transfer records from the Allbridge Core Explorer backend.
RPC verification, transaction decoding, deduplication, and
unstable-target filtering yield 1,341 valid pairs, including
795 from Ethereum and 182 from each of Arbitrum, Base, and
Polygon PoS.

\item \textbf{deBridge:}
We use deBridge DLN as the representative SFT product and collect
1,600 fulfilled orders from the official Stats API. RPC-based
verification of source orders and Solana fulfillment transactions
yields 1,597 valid source--target pairs, including 997 from
Ethereum and 200 from each of Arbitrum, Base, and Polygon PoS.

\item \textbf{Circle:}
We use Circle CCTP as the representative BMT product and index
29,260 source-side burn records using Dune Analytics, Circle Iris,
and CCTP program histories. After source-chain filtering,
message-level deduplication, and semantic consistency verification
against Solana receive-and-mint transactions, 2,519 valid
burn--mint pairs are retained, including 1,495 from Ethereum,
385 from Arbitrum, 477 from Base, and 162 from Polygon PoS.
\end{itemize}

As summarized in Table~\ref{tab:bridge_datasets}, each verified
source--target association is treated as a correlated pair (CP).
For each CP, we construct 30 protocol-compatible non-correlated
pairs (NCPs) by pairing the source transaction with distractor
transactions on Solana. This yields 212,460 NCPs and 219,542
candidate pairs in total. The \emph{Raw Data} column reports the
number of records obtained during data acquisition, while the
per-source-chain CP/NCP columns report the verified associations
and their corresponding constructed distractors.

\subsection{Experimental Settings}

\textit{1) Environments:} All experiments are conducted on a laptop running Windows 11 64-bit, equipped with an AMD Ryzen 9 5900HX 8-Core CPU, an NVIDIA GeForce RTX 3060 Laptop GPU with 6 GB memory, and 16 GB RAM. The experimental programs are implemented in Python 3.13.10.

\textit{2) Methods in Comparison:} To comprehensively evaluate the performance of SolTracer, we employ six typical methods for comparison, which can be categorized into two categorizes.

Heuristic-based methods:
	\begin{itemize}
		\item Attribute Matching (AttrMatch)~\cite{yousaf2019tracing}  scores candidate transaction pairs based on time, amount, and asset consistency, accepting associations above a tuned threshold.
		
		\item Entity-Aware Matching (EntityMatch)~\cite{yousaf2019tracing} extends the AttrMatch by incorporating recipient and counterparty consistency into the threshold-based association decision.
		
		\item Multi-Constraint Matching (MultiMatch)~\cite{yan2025evmpolygon} establishes a deterministic rule-based framework over reconstructed events, associating cross-chain transactions only when a candidate uniquely satisfies strict account, timing, amount, and token constraints.
	\end{itemize}
	
Learning-based methods:
		\begin{itemize}
		\item CONNECTOR~\cite{lin2025connector} extracts bridge-contract execution features to identify deposit events on the source chain and links them to corresponding target withdrawals using on-chain execution evidence.
		
		\item ABCTRACER~\cite{lin2025tracktrace} combines event mining and named-entity recognition for explicit cross-chain cue extraction with an information-retrieval ranking model for implicit cue matching.
		
		\item ConneX~\cite{liang2025connex} leverages large language models to extract semantic key-value pairs from transaction logs and applies programmatic value-consistency checks to identify valid target evidence.
	\end{itemize}

To adapt these methods to the heterogeneous EVM-to-Solana setting,
we replace their EVM-specific input fields with a unified semantic-event
representation.AttrMatch, EntityMatch, and MultiMatch apply their original scoring or
constraint rules to the corresponding time, amount, asset, and
participant fields. For CONNECTOR, ABCTRACER, and
ConneX, EVM-specific log inputs are replaced with the
corresponding execution evidence or semantic key--value records, while
their original association mechanisms are retained.

\textit{3) Metric:} We evaluate performance using precision  ($P$), recall  ($R$), and F1-score  ($F1$) as they are standard in this research field \cite{SolPhish, FiMAD, wu2025safeguarding}. In open-world scenarios with selective rejection ~\cite{geifman2019selectivenet}, group-level metrics ($P_g, R_g, F1_g$) quantify the capability to abstain when no valid target exists:

\begin{equation}
	P_g = \frac{\text{TP}_g}{\text{TP}_g + \text{FP}_g}, \quad
	R_g = \frac{\text{TP}_g}{\text{TP}_g + \text{FN}_g}, \quad
	F1_g = \frac{2 P_g R_g}{P_g + R_g}
	\label{eq:evaluation-group-metrics}
\end{equation}
where $\text{TP}_g$, $\text{FP}_g$, and $\text{FN}_g$ denote correctly rejected target-absent groups, falsely rejected target-present groups, and unrejected target-absent groups, respectively.

\textit{4) Validation.} A candidate group is defined as Target-Present
(TP) when it contains a verified Solana-side target, and as
Target-Absent (TA) when the verified target is removed through
controlled masking. We vary the TP/TA ratio to evaluate both
closed-world association and open-world association under different
degrees of class imbalance. All datasets are partitioned at the
candidate-group level into mutually exclusive training, validation,
and test sets. The closed- and open-world experiments use
Ethereum-to-Solana data, whereas cross-source-chain generalization
uses Ethereum for training and calibration and Arbitrum, Base, and
Polygon PoS for evaluation. 

\textit{5) Implementation Details.} During candidate ranking,
\textsc{SolTracer} employs a random forest over 14 semantic features,
with 300 trees, balanced class weights, and a minimum leaf size of two.
The candidate-set rejector uses ranking and semantic-consistency evidence,
while excluding candidate count and minimum time distance to prevent
structural leakage. To reduce the impact of randomness,the main closed-world, open-world, and generalization experiments 
are each repeated with five fixed seeds (7, 11, 19, 23, and 42), and the mean results are reported.

\subsection{Sensitivity Analysis}\label{sec:threshold_sensitivity}
\begin{figure}[h]
	\centering
	\includegraphics[width=0.95\columnwidth]{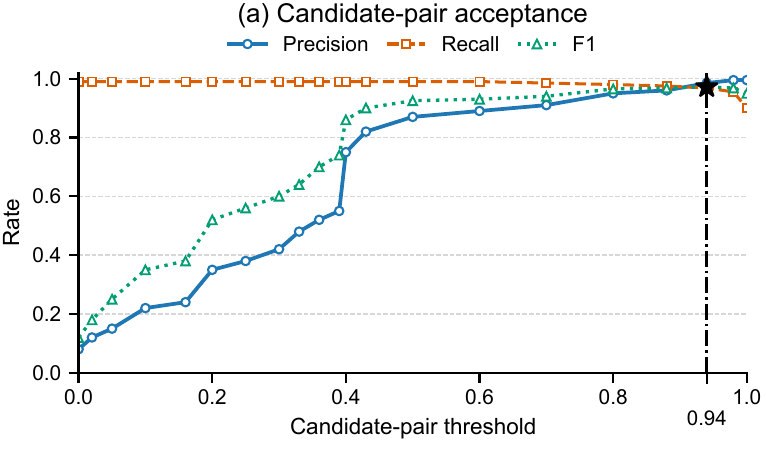}
	\caption{Sensitivity of candidate-pair acceptance across thresholds $\tau_p \in [0, 1]$. The star and vertical dashed line indicate the optimal threshold ($\tau_p = 0.94$) maximizing F1.}
	\label{fig:sensitivity-pair}
\end{figure}

\textit{1) Candidate-pair threshold $\tau_p$}. The threshold $\tau_p$ determines whether an individual cross-chain association is accepted in closed-world scenarios. To find the optimal operating point without data leakage, we perform a grid sweep of $\tau_p$ from 0 to 1 with a step size of 0.01 over the 12,400 validation pairs (371 positive and 12,029 negative) to maximize the candidate-level F1-score. As shown in Fig.~\ref{fig:sensitivity-pair}(a), low thresholds retain nearly all true targets but admit substantial negative distractors, leading to low precision. As $\tau_p$ increases, precision improves monotonically while recall remains stable above 0.92 until $\tau_p = 0.98$. At $\tau_p = 0.94$, candidate F1 peaks at 97.25\%. Notably, the F1 curve exhibits a stable plateau across $[90\%, 98\%]$, demonstrating SolTracer's robustness against threshold variations over an isolated peak.

\begin{figure}[h]
	\centering
	\includegraphics[width=0.95\columnwidth]{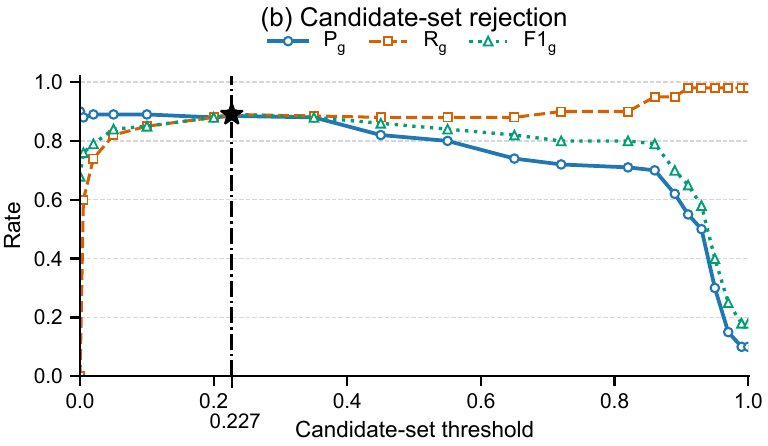}
	\caption{Sensitivity of candidate-set rejection across thresholds $\tau_r \in [0, 1]$. The star and vertical dashed line indicate the optimal operating point ($\tau_r = 0.227$) maximizing group-level F1.}
	\label{fig:sensitivity-set}
\end{figure}

\begin{table*}[!t]
	\caption{Closed-world association performance (\%) across different transaction modes.}
	\label{tab:closed_world_mode_comparison}
	\centering
	\footnotesize
	\renewcommand{\arraystretch}{1.15}
	\setlength{\tabcolsep}{3.8pt}
	\sisetup{
		table-number-alignment = center,
		table-format = 2.2,
		detect-weight = true
	}
	\begin{tabular}{
		@{}l
		*{4}{
			S[table-format=2.2]
			S[table-format=2.2]
			S[table-format=2.2]
		}
		S[table-format=2.2]
		S[table-format=2.2]
		S[table-format=2.2]@{}
	}
		\toprule
		\multirow{2}{*}{\textbf{Method}}
		& \multicolumn{3}{c}{MRT}
		& \multicolumn{3}{c}{PST}
		& \multicolumn{3}{c}{SFT}
		& \multicolumn{3}{c}{BMT}
		& \multicolumn{3}{c}{\textbf{Overall}} \\
		
		\cmidrule(lr){2-4}
		\cmidrule(lr){5-7}
		\cmidrule(lr){8-10}
		\cmidrule(lr){11-13}
		\cmidrule(lr){14-16}
		
		& {$P$} & {$R$} & {$F1$}
		& {$P$} & {$R$} & {$F1$}
		& {$P$} & {$R$} & {$F1$}
		& {$P$} & {$R$} & {$F1$}
		& {$P$} & {$R$} & {$F1$} \\
		\midrule
		
		AttrMatch
		& 0.00 & 0.00 & 0.00
		& 97.83 & 100.00 & 98.90
		& 85.05 & 82.73 & 83.87
		& 80.91 & 98.89 & 89.00
		& 87.38 & 71.05 & 78.37 \\
		
		EntityMatch
		& 0.00 & 0.00 & 0.00
		& 97.83 & 100.00 & 98.90
		& 79.82 & 82.73 & 81.25
		& 78.07 & 98.89 & 87.25
		& 84.38 & 71.05 & 77.14 \\
		
		MultiMatch
		& 100.00 & 93.33 & 96.55
		& 0.00 & 0.00 & 0.00
		& 0.00 & 0.00 & 0.00
		& 100.00 & 78.89 & 88.20
		& 100.00 & 40.79 & 57.94 \\
		
		CONNECTOR
		& 0.00 & 0.00 & 0.00
		& 96.77 & 100.00 & 98.36
		& 81.82 & 65.45 & 72.73
		& 75.42 & 98.89 & 85.58
		& 83.95 & 66.05 & 73.93 \\
		
		ABCTRACER
		& 86.30 & 86.30 & 86.30
		& 85.19 & 85.19 & 85.19
		& 70.74 & 57.88 & 63.67
		& 92.96 & 92.96 & 92.96
		& 83.80 & 79.39 & 81.53 \\
		
		ConneX
		& 88.76 & 87.78 & 88.27
		& 97.78 & 97.78 & 97.78
		& 84.71 & 65.45 & 73.85
		& 84.44 & 84.44 & 84.44
		& 88.98 & 82.89 & 85.83 \\
		
		\midrule
		\textbf{SolTracer}
		& \textbf{100.00} & \textbf{96.67} & \textbf{98.31}
		& \textbf{98.82} & 93.33 & 96.00
		& \textbf{99.10} & \textbf{100.00} & \textbf{99.55}
		& \textbf{100.00} & 96.67 & \textbf{98.31}
		& 99.46 & \textbf{96.84} & \textbf{98.13} \\
		
		\bottomrule
	\end{tabular}
\end{table*}

\textit{2) Candidate-set rejection threshold $\tau_r$}. The threshold $\tau_r$ dictates whether a ranked candidate pool provides sufficient evidence to return any association in open-world scenarios. We generate out-of-fold support probabilities via 5-fold stratified cross-validation on the 400 validation groups, selecting the threshold $\tau_r$ that maximizes the $F1_g$. As shown in Fig.~\ref{fig:sensitivity-set}(b), the optimal boundary is identified at $\tau_r = 0.227$, achieving $P_g = 89.74\%$, $R_g = 87.50\%$, and $F1_g = 88.61\%$ by correctly rejecting 35 of the 40 TA validation groups while incorrectly rejecting only 4 of the 360 TP groups. Beyond this operating point, precision degrades sharply because the model aggressively rejects valid TP groups despite higher TA recall. These two complementary curves confirm that pair-level filtering and set-level abstention govern distinct error modes and achieve stable performance when calibrated independently.

\subsection{Evaluation in Closed-World Scenario}

The closed-world settings evaluates candidate discrimination when the ground-truth target is guaranteed to exist within the observation pool across diverse bridge architectures. We partition 2,340 positive groups into 1,620 training, 360 validation, and 360 held-out test groups with 90 groups per transaction mode. Table~\ref{tab:closed_world_mode_comparison} reports candidate-level F1-scores across all four individual transaction modes along with overall performance on the complete 360-group test set. We have the following main observations.

1) SolTracer achieves the best overall association
performance across heterogeneous transaction modes.
SolTracer attains an overall F1 of 98.13\%, outperforming the strongest
baseline, ConneX, by 12.30 percentage points, while maintaining both
high precision (99.46\%) and recall (96.84\%). In contrast, existing
methods exhibit strong mode dependence: strict constraints cause
MultiMatch to miss valid targets, whereas attribute- and execution-based
methods fail when their matching cues are not preserved across mechanisms.
This demonstrates that mode-guided semantic reconstruction provides more
consistent association evidence across heterogeneous bridge transactions.

2) SolTracer shows greater advantages under stronger
semantic transformations.
SolTracer achieves the highest F1 on MRT, SFT, and BMT, reaching
98.31\%, 99.55\%, and 98.31\%, respectively, with the largest gain
of 15.68 percentage points on SFT. This advantage stems from recovering
mechanism-specific evidence such as redemption roles, solver fulfillment,
and burn--mint semantics. On PST, AttrMatch and EntityMatch reach
98.90\%, slightly above SolTracer's 96.00\%, because time, asset, and
amount consistency remains sufficiently discriminative in this mode.

\begin{table}[!t]
\caption{open-world association performance (\%) across different methods.}
\label{tab:open_world_comparison}
\centering
\footnotesize
\renewcommand{\arraystretch}{1.15}
\setlength{\tabcolsep}{3.5pt}
\sisetup{
    table-number-alignment = center,
    table-format = 2.2,
    detect-weight = true,
    detect-inline-weight = math
}
\begin{tabular}{
    @{}l
    *{6}{S[table-format=2.2]}
    @{}
}
\toprule
\multirow{2}{*}{\textbf{Method}}
& \multicolumn{2}{c}{10\%}
& \multicolumn{2}{c}{30\%}
& \multicolumn{2}{c}{50\%} \\
\cmidrule(lr){2-3}
\cmidrule(lr){4-5}
\cmidrule(l){6-7}
& {$F1$} & {$F1_g$}
& {$F1$} & {$F1_g$}
& {$F1$} & {$F1_g$} \\
\midrule
AttrMatch
& 78.03 & 44.05
& 76.36 & 72.67
& 73.71 & 84.60 \\

EntityMatch
& 76.70 & 44.05
& 74.73 & 71.81
& 71.60 & 83.80 \\

MultiMatch
& 57.94 & 28.07
& 57.42 & 60.00
& 55.70 & 77.37 \\

CONNECTOR
& 72.75 & 38.46
& 70.65 & 68.18
& 67.31 & 81.28 \\

ABCTRACER
& 77.35 & {--}
& 67.81 & {--}
& 55.07 & {--} \\

ConneX
& 83.33 & 56.25
& 77.92 & 65.22
& 69.25 & 63.76 \\

\midrule
\textbf{SolTracer}
& \bfseries 96.50 & \bfseries 94.87
& \bfseries 95.22 & \bfseries 95.69
& \bfseries 93.87 & \bfseries 97.19 \\
\bottomrule
\end{tabular}
\end{table}

\begin{table*}[!t]
	\centering
	\caption{Cross-source-chain generalization performance (\%) across different methods.}
	\label{tab:generalization}
	\setlength{\tabcolsep}{4pt}
	\renewcommand{\arraystretch}{1.08}
	\resizebox{\textwidth}{!}{
		\begin{tabular}{lcccccccccccccccc}
			\toprule
			\multirow{2}{*}{Method}
			& \multicolumn{4}{c}{Arbitrum-to-Solana}
			& \multicolumn{4}{c}{Base-to-Solana}
			& \multicolumn{4}{c}{PolygonPoS-to-Solana}
			& \multicolumn{4}{c}{Overall} \\
			\cmidrule(lr){2-5}
			\cmidrule(lr){6-9}
			\cmidrule(lr){10-13}
			\cmidrule(l){14-17}
			& $P$ & $R$ & $F1$ & $F1_g$
			& $P$ & $R$ & $F1$ & $F1_g$
			& $P$ & $R$ & $F1$ & $F1_g$
			& $P$ & $R$ & $F1$ & $F1_g$ \\
			\midrule
			
			AttrMatch
			& 81.50 & 94.44 & 83.79 & 62.50
			& 83.41 & 86.11 & 82.97 & 82.14
			& 86.70 & 100.00 & 92.61 & 100.00
			& 83.87 & 93.52 & 86.46 & 81.55 \\
			
			EntityMatch
			& 74.91 & 100.00 & 83.24 & 50.00
			& 79.01 & 100.00 & 88.07 & 100.00
			& 78.27 & 100.00 & 86.93 & 75.00
			& 77.40 & 100.00 & 86.08 & 75.00 \\
			
			MultiMatch
			& 75.00 & 72.22 & 73.53 & 71.21
			& 75.00 & 75.00 & 75.00 & 79.55
			& 75.00 & 75.00 & 75.00 & 79.55
			& 75.00 & 74.07 & 74.51 & 76.77 \\
			
			CONNECTOR
			& 68.66 & 100.00 & 79.66 & 50.00
			& 77.31 & 100.00 & 87.00 & 100.00
			& 75.77 & 100.00 & 85.61 & 75.00
			& 73.91 & 100.00 & 84.09 & 75.00 \\
			
			ABCTRACER
			& 37.50 & 41.67 & 39.47 & 0.00
			& 38.33 & 42.59 & 40.35 & 0.00
			& 45.83 & 50.93 & 48.25 & 0.00
			& 40.55 & 45.06 & 42.69 & 0.00 \\
			
			ConneX
			& 77.50 & 86.11 & 81.58 & 50.00
			& 79.44 & 83.33 & 81.29 & 50.00
			& 90.00 & 97.22 & 93.42 & 25.00
			& 82.31 & 88.89 & 85.43 & 41.67 \\
			
			\midrule
			\textbf{SolTracer}
			& \textbf{100.00} & 93.89 & \textbf{96.06} & \textbf{89.76}
			& \textbf{100.00} & 97.78 & \textbf{98.79} & 94.17
			& \textbf{100.00} & 97.22 & \textbf{98.53} & 91.67
			& \textbf{100.00} & 96.30 & \textbf{97.79} & \textbf{91.87} \\
			
			\bottomrule
		\end{tabular}
	}
\end{table*}

\subsection{Evaluation in Open-World Scenario}

The open-world protocol relaxes the target-availability assumption to
evaluate whether a method can selectively abstain when valid target
evidence is unavailable~\cite{geifman2019selectivenet}. We use the
same 2,600 Ethereum-to-Solana groups, partitioned into 1,800 training,
400 validation, and 400 test groups, and construct three open-world
settings with TA ratios of 10\%, 30\%, and 50\%. For each setting,
the corresponding proportion of groups is designated as Target-Absent
by removing the verified target while retaining protocol-compatible
distractors. Table~\ref{tab:open_world_comparison}
reports  $F1$ and $F1_g$ under the three TA ratios. We have the following observations.

1) Increasing target absence exposes the limitation
of pair-level association. As the TA ratio increases, the candidate-level performance of all
baseline methods generally deteriorates. For example, ConneX decreases
from 83.33\% to 69.25\% in $F1$, while ABCTRACER drops more sharply
from 77.35\% to 55.07\%. This degradation indicates that methods
designed primarily to identify the best candidate become increasingly
vulnerable when valid targets are absent from a larger fraction of
candidate sets. Although several heuristic methods obtain higher
$F1_g$ at larger TA ratios, their candidate-level $F1$ continues to
decline, showing that stronger rejection under frequent target absence
does not necessarily imply reliable association of target-present
groups.

2) Candidate-set selective decision remains robust
across different TA ratios. SolTracer consistently achieves the highest candidate-level $F1$,
decreasing only moderately from 96.50\% at 10\% TA to 93.87\% at
50\% TA. In contrast, its rejection $F1_g$ remains above 94\% across
all settings and increases from 94.87\% to 97.19\% as target absence
becomes more prevalent. Moreover, its candidate-level advantage over
the strongest baseline widens from 13.17 percentage points at 10\% TA
to 20.16 percentage points at 50\% TA. These results show that
separating within-set ranking from candidate-set-level abstention
allows SolTracer to preserve accurate target association while
adapting to substantial variation in target availability, making it
more reliable under the imbalanced conditions encountered in
real-world cross-chain tracing.

\begin{table}[!t]
	\caption{Ablation results on individual SolTracer modules (\%).}
	\label{tab:unified-ablation}
	\centering
	\footnotesize
	\setlength{\tabcolsep}{2.0pt}
	\renewcommand{\arraystretch}{1.15}
	\begin{tabular*}{\columnwidth}{@{\extracolsep{\fill}}llcc@{}}
		\toprule
		\textbf{Category} & \textbf{Configuration} & \textbf{$F1$} & \textbf{$F1_g$} \\
		\midrule
		\multirow{2}{*}{\shortstack[l]{Protocol\\Identification}} 
		& Protocol-Guided$^\ast$ & \textbf{96.50} & \textbf{94.87} \\
		& w/o Protocol-Guided & 75.57 \textcolor{red}{\tiny $\downarrow$20.93} & 37.86 \textcolor{red}{\tiny $\downarrow$57.01} \\
		\midrule
		\multirow{2}{*}{\shortstack[l]{Event\\Reconstruction}} 
		& Mode-Guided$^\ast$ & \textbf{96.90} & \textbf{98.73} \\
		& w/ Generic Repr. & 96.37 \textcolor{red}{\tiny $\downarrow$0.53} & 96.10 \textcolor{red}{\tiny $\downarrow$2.63} \\
		\midrule
		\multirow{3}{*}{\shortstack[l]{Classifier}} 
		& Random Forest$^\ast$ & 96.50 & \textbf{94.87} \\
		& w/ Logistic Regression & 77.98 \textcolor{red}{\tiny $\downarrow$18.52} & 62.92 \textcolor{red}{\tiny $\downarrow$31.95} \\
		& w/ XGBoost & 96.50 & 91.89 \textcolor{red}{\tiny $\downarrow$2.98} \\
		\midrule
		\multirow{2}{*}{\shortstack[l]{Association\\Decision}} 
		& Selective Decision$^\ast$ & \textbf{96.50} & \textbf{94.87} \\
		& w/o Selective Decision & 92.05 \textcolor{red}{\tiny $\downarrow$4.45} & -- \\
		\bottomrule
		\multicolumn{4}{@{}l}{\scriptsize $^\ast$ Default setting. Red values ($\downarrow$) denote performance degradation from the default.}
	\end{tabular*}
    \vspace{-1.2\baselineskip}
\end{table}

\subsection{Evaluation on Generalization}

To evaluate cross-source-chain generalization, we calibrate all methods on Ethereum-to-Solana transfers and directly apply the frozen configurations to Arbitrum, Base, and Polygon PoS. Table~\ref{tab:generalization} reports candidate association and candidate-set rejection performance on the three unseen source chains. We have the following observations.

1) SolTracer achieves the strongest overall cross-source-chain generalization.
SolTracer obtains an overall candidate F1 of 97.79\% and rejection
$F1_g$ of 91.87\%, exceeding the strongest baselines by 11.33 and
10.32 percentage points, respectively. In contrast, heuristic methods
remain dependent on predefined attribute assumptions, while learned
baselines degrade when source-chain-specific execution patterns shift.
The results indicate that protocol- and mode-guided semantic evidence
provides a more transferable representation across EVM source chains.

2) SolTracer remains consistently effective across
all unseen source chains.
Its candidate F1 reaches 96.06\%, 98.79\%, and 98.53\% on Arbitrum,
Base, and Polygon PoS, outperforming the strongest baseline by
12.27, 10.72, and 5.11 percentage points, respectively. Although
ConneX becomes more competitive on Polygon PoS, other baselines show
larger variations across source chains. SolTracer's consistently high
association and rejection performance suggests that semantic
reconstruction reduces sensitivity to source-chain distribution shifts.

\subsection{Ablation Study}

In this subsection, we systematically evaluate the contribution of each core module in SolTracer, including the protocol identification module, the semantic event reconstruction module, the candidate classifier, and the candidate-set selective decision mechanism. All variants are evaluated on the standardized test benchmark containing 360 target-present and 40 target-absent groups. Table~\ref{tab:unified-ablation} presents the association and rejection performance across different configurations.

\par\indent 1) Protocol Identification. Without protocol-guided filtering, the candidate matching F1 drops sharply by 20.93\%, and the rejection score $F1_g$ collapses from 94.87\% to 37.86\%. This severe performance degradation occurs because generic matching relies strictly on raw attribute consistency, which fails under cross-chain conversions and causes supported targets to be incorrectly filtered out.

\par\indent 2) Event Reconstruction. Replacing the mode-guided reconstruction with a generic representation reduces the candidate F1 to 96.37\% and causes a 2.63\% drop in set-level rejection $F1_g$. Reconstructing mode-specific account roles and inner invocations provides indispensable semantic cues for distinguishing target candidates and assessing candidate-set support.

\par\indent 3) Classifier. Linear models like Logistic Regression achieve an F1 of only 77.98\% and an $F1_g$ of 62.92\%, highlighting the need for nonlinear modeling across heterogeneous feature interactions. While XGBoost achieves a comparable candidate F1 of 96.50\%, its set rejection $F1_g$ degrades by 2.98\%, confirming that Random Forest delivers the most balanced rejection capability.

\par\indent 4) Association Decision. Removing the candidate-set reject option and forcing the model to return the top candidate decreases candidate F1 from 96.50\% to 92.05\%, while completely losing the ability to identify target-absent groups. This confirms that selective abstention is crucial to prevent forced false-positive attributions in open-world tracing.

\section{Empirical Insights on Solana Cross-Chain Transactions}\label{sec: EmpiricalStudy}

In this section, we analyze the cross-chain transactions associated by \textsc{SolTracer} to uncover key operational and security characteristics of the Solana bridge ecosystem. Specifically, we investigate count--value divergence, cross-asset shifts, target observability gaps, and mechanism-dependent association reliability.

\subsection{Difference Between Transaction Count and Value}

\begin{figure}[!htbp]
    \centering
    \includegraphics[width=0.82\columnwidth]
    {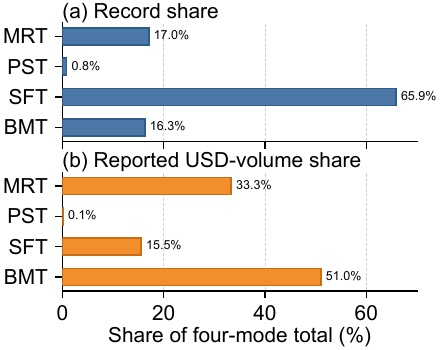}
    \caption{Record and reported USD-volume shares across the four Ethereum-to-Solana transaction modes.}
    \label{fig:protocol-activity}
\end{figure}

To examine whether transaction frequency reflects the economic concentration of Solana-bound bridge activity, we compare the record share against the reported USD-volume share across the four transaction modes. We restrict this analysis to Ethereum-to-Solana transfers collected over a 28-day window from March 25 to April 22, 2026 UTC, spanning 16,790 records and \$178.53 million in source-side volume. Fig.~\ref{fig:protocol-activity} illustrates the distribution.

As shown in Fig.~\ref{fig:protocol-activity}, SFT accounts for 65.86\% of total transactions but only 15.53\% of volume. Its intent-based solver model predominantly handles high-frequency, price-sensitive DEX swaps and cross-chain arbitrage of relatively small amounts~\cite{debridge_dln_overview,debridge_swaps}. Conversely, BMT exhibits the opposite pattern, contributing only 16.28\% of transactions while commanding 51.03\% of total volume. Its native burn-and-mint mechanism serves large-scale capital deployments and liquidity rebalancing by market makers and institutional entities~\cite{circle_cctp_v1}.

Similarly, MRT shows a higher volume share than transaction share, reflecting direct high-value asset transfers. PST accounts for minimal activity across both metrics on the evaluated route. Overall, transaction frequency does not directly reflect economic exposure: frequent solver orders process fragmented retail volume, whereas less frequent burn-and-mint operations concentrate substantial liquidity.

\findingbox{\textbf{Finding 1.} Solana-bound bridge activity exhibits a stark count--value divergence. SFT generates the highest transaction count, while BMT and MRT account for the vast majority of transferred value. Monitoring frameworks relying solely on transaction frequency fail to capture high-value systemic capital flows.}

\subsection{Cross-Asset Settlement Patterns}

\begin{figure}[!htbp]
	\centering
	\includegraphics[width=\columnwidth]{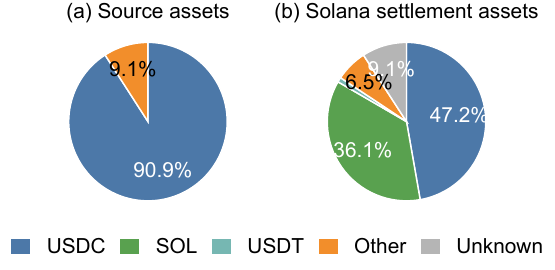}
	\caption{Source-side and Solana-side asset composition in the indexed dataset. \emph{Unknown} indicates missing target-side token symbols.}
	\label{fig:asset-composition}
\end{figure}

To examine how intermediate token conversions impact fund lineage across heterogeneous chains, we analyze the distribution shift between source-side deposit assets and Solana-side settlement assets across all indexed transfers. Specifically, we quantify the token breakdown across both chains and measure the degree of asset transformation induced by automated DEX routing and solver fulfillment. Fig.~\ref{fig:asset-composition} illustrates the compositional divergence across the 47,117 collected cross-chain records.

As depicted in the distributions, USDC dominates source-chain deposits at 90.90\%, but its representation plunges to 47.24\% upon Solana settlement. Concurrently, native SOL comprises 36.13\% of settlement assets, while unspecified tokens, alternative SPL assets, and USDT account for 9.13\%, 6.51\%, and 0.98\%, respectively. Overall, non-USDC assets surge from 9.10\% on the source chain to 52.76\% on Solana, demonstrating that cross-asset settlement is the prevailing pattern rather than an exception.

While cross-asset swaps offer flexibility for retail users, they inherently disrupt the semantic continuity of cross-chain fund flows. By altering token contracts, payout values, and decimal precisions within solver executions, cross-asset routes allow illicit actors to mimic legitimate DEX swaps and bypass static AML amount-matching rules. Tracing frameworks that enforce raw asset or balance equality inevitably fail across such intent-based mechanisms.

\findingbox{\textbf{Finding 2.} Cross-chain settlements frequently transform asset composition, with non-USDC shares surging from 9.10\% to 52.76\% on Solana. Because automated cross-asset swaps alter fundamental transfer attributes, reliable tracing cannot assume source--target asset equality.}

\subsection{Separation Between Settlement and Visibility}

\begin{figure}[!htbp]
    \centering
    \includegraphics[width=0.88\columnwidth]
    {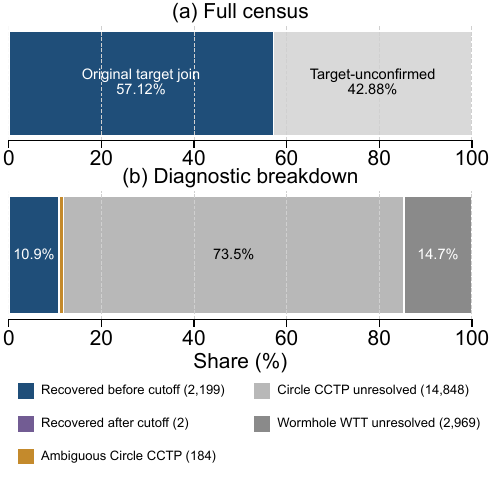}
    \caption{Distribution of source-only records and the empirical
    breakdown resolved by \textsc{SolTracer}.}
    \label{fig:unconfirmed-diagnostics}
\end{figure}

To assess whether missing target joins in explorer records stem from actual cross-chain execution failures or target observability gaps on Solana, we investigate the resolution status of all unlinked source-side transactions. Specifically, we replay these records through \textsc{SolTracer} against bounded Solana candidate pools and mode-specific semantic evidence to differentiate pre-cutoff settlements, candidate ambiguity, and genuinely unredeemed transfers. Fig.~\ref{fig:unconfirmed-diagnostics} presents the overall proportion of source-only records and the empirical breakdown of recovered targets.

Among the 47,117 indexed bridge records, 20,202 transfers (42.88\%) contain only an origin deposit without an explicit target-side link. Our replay analysis resolves these unlinked transfers into three distinct categories:

\textit{1) Settled but Unlinked Transfers.} \textsc{SolTracer} recovers 2,201 unique target transactions from the source-only subset. Crucially, 2,199 of these targets (1,493 Circle CCTP and 706 Wormhole WTT transactions) had settled successfully on-chain prior to the collection cutoff, while only two settled afterward. This confirms that missing links primarily reflect target-side indexing omissions rather than long settlement delays. Because Solana scatters execution state across inner CPIs and balance deltas without emitting standard top-level receipt logs, general-purpose explorers fail to capture and associate these completed payouts.

\textit{2) Ambiguous Associations.} Another 184 Circle CCTP transfers match multiple concurrent mint operations satisfying identical temporal, recipient, and amount constraints. In such cases, observable target evidence exists but cannot support a unique attribution, highlighting the necessity of set-level selective rejection to avoid false linkages.

\textit{3) Unresolved Transfers.} The remaining transfers (14,848 Circle CCTP and 2,969 Wormhole WTT records) yield no valid match within the observation window. These records correspond to unsupported custom routes, long-tail finality delays, or genuinely unexecuted transactions. Rather than forcing false associations, \textsc{SolTracer} abstains from linking them.

\findingbox{\textbf{Finding 3.} On-chain settlement on Solana is decoupled from receipt-level observability. Over 10.8\% of unindexed records had settled successfully on-chain prior to the collection cutoff, demonstrating that standard explorers miss valid transactions due to CPI-fragmented execution traces.}

\subsection{Mode-Dependent Association Reliability}

\begin{figure}[!htbp]
    \centering
    \includegraphics[width=0.80\columnwidth]
    {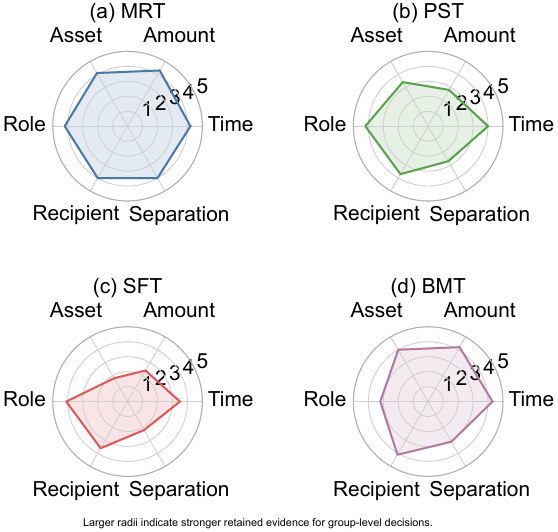}
    \caption{Discriminative evidence profiles across the four Solana
    cross-chain transaction modes.}
    \label{fig:mode-evidence-profile}
\end{figure}

To understand how disparate bridge architectures influence cross-chain tracing reliability under open-world target uncertainty, we examine the discriminative strength of candidate-set evidence across the four transaction modes. Specifically, we evaluate the error distributions and candidate separation profiles across time, amount, asset, recipient, and role cues under both target-present and target-absent settings. Fig.~\ref{fig:mode-evidence-profile} summarizes the mechanism-specific evidence retention patterns for candidate-set decisions.

The empirical evidence profiles explain the systematic divergence in association and rejection performance across modes:

MRT preserves direct recipient and redemption-role cues once inner CPI invocations to the SPL Token program are resolved. PST introduces amount deviations through dynamic pool slippage and liquidity fees, while shared pool contracts generate concurrent payouts with overlapping value ranges. SFT exhibits the weakest direct linkage because solver fulfillment decouples funding accounts and permits cross-asset conversions. BMT maintains clean recipient and amount semantics, yet its off-chain attestation verification makes candidate separation vulnerable to concurrent identical-value mint batches.

\findingbox{\textbf{Finding 4.} Association and rejection reliability depends heavily on the underlying bridge mechanism. Because evidence retention profiles differ structurally across MRT, PST, SFT, and BMT, reliable open-world tracing requires mode-specific feature modeling and calibrated abstention.}

\section{Conclusion}\label{sec: conclusion}

This work is the first to systematically explore cross-chain transaction tracing on Solana across heterogeneous ledgers. We analyzed the unique execution models and transaction record designs between EVM and Solana, identifying four fundamental Solana-bound transaction modes based on their settlement mechanisms, namely MRT, PST, SFT, and BMT. Subsequently, we proposed SolTracer, a cross-chain tracing method based on candidate-set selective decision, which effectively reconstructs semantic bridge events and reliably associates target transactions.Extensive experiments demonstrate that SolTracer outperforms state-of-the-art methods across closed-world, open-world, and cross-source-chain generalization scenarios. In the open-world scenario, SolTracer improves the F1 score by up to 20.16\% over the strongest baseline. We then conducted an empirical analysis on real-world cross-chain transfers, exploring the count-value divergence, cross-asset settlement shifts, target-observability gaps, and mechanism-dependent association reliability. In the future, we will extend method identification and selective decision-making to evolving routes and multi-hop cross-chain settlements.

\bibliographystyle{IEEEtran}
\bibliography{references}

\end{document}